# Non-Invasive Detection of **PROS**tate Cancer with Novel **T**ime-**D**ependent Diffusion MRI and **AI**-Enhanced Quantitative Radiological Interpretation: PROS-TD-AI

**Baltasar Ramos [1], Cristian Garrido [1], Paulette Narváez [2], Santiago Gelerstein Claro [3], Haotian Li [4], Rafael Salvador [5], Constanza Vásquez-Venegas [6], Iván Gallegos [7], Yi Zhang [4], Víctor Castañeda [8], Cristian Acevedo [9], Dan Wu [4], Gonzalo Cárdenas [1] and Camilo G. Sotomayor [1,10,11] \***

[1] Radiology Department, Clinical Hospital of the University of Chile, University of Chile, Independencia 8380453, Chile; baltasarramos@ug.uchile.cl, cgarrido@hcuch.cl, gcardenas@hcuch.cl, camilosotomayor@uchile.cl

[2] Urology Department, Clínica Dávila, Santiago 8431657, Chile; kikinarvaez@gmail.com

[3] School of Medicine, Faculty of Medicine, University of Chile, Independencia 8380453, Santiago, Chile; santiago.gelerstein@ug.uchile.cl

[4] Key Laboratory for Biomedical Engineering of Ministry of Education, College of Biomedical Engineering & Instrument Science, Zhejiang University, Hangzhou, China; haotianli@zju.edu.cn, yizhangzju@zju.edu.cn, danwu.bme@zju.edu.cn

[5] Radiology Department, Hospital Clinic, Universitat de Barcelona, 170, 08036, Barcelona, Spain; rsalvadoriz@gmail.com

[6] Laboratory for Scientific Image Analysis SCIAN-Lab, Integrative Biology Program, Institute of Biomedical Sciences ICBM, Faculty of Medicine, University of Chile, Av. Independencia 1027, Santiago 8380453, Chile; covasquezv@inf.udec.cl

[7] Pathology Department, Clinical Hospital of the University of Chile, University of Chile, Independencia 8380453, Chile; igallegosmendez@gmail.com

[8] Medical Technology Department, Faculty of Medicine, University of Chile, Santiago 8380453, Chile; vcastane@uchile.cl

[9] Urology Department, Clinical Hospital of the University of Chile, University of Chile, Independencia 8380453, Chile; cacevedo@hcuch.cl

[10] Anatomy and Developmental Biology Program, Institute of Biomedical Sciences, Faculty of Medicine, University of Chile, Santiago 8380453, Chile; camilosotomayor@uchile.cl

[11] Faculty of Medicine, San Sebastián University, Campus Los Leones, Lota 2465, Providencia 7510157, Santiago, Chile; camilo.sotomayor@uss.cl

\* Correspondence: Camilo G. Sotomayor, MD, PhD. E-mail: camilosotomayor@uchile.cl; Tel. 562 2978 8412

**Abstract**

Prostate cancer (PCa) is the most frequently diagnosed malignancy in men and the eighth leading cause of cancer death worldwide. Multiparametric MRI (mpMRI) has become central to the diagnostic pathway for men at intermediate risk, improving detection of clinically significant PCa (csPCa) while reducing unnecessary biopsies and over-diagnosis. However, mpMRI remains limited by false positives, false negatives, and moderate to substantial interobserver agreement. Time-dependent diffusion (TDD) MRI, a novel sequence that enables tissue microstructure characterization, has shown encouraging preclinical performance in distinguishing clinically significant from insignificant PCa. Combining TDD-derived metrics with machine learning may provide robust, zone-specific risk prediction with less dependence on reader training and improved accuracy compared to current standard-of-care. This *study protocol* outlines the rationale and describes the prospective evaluation of a home-developed AI-enhanced TDD-MRI software (PROST-DAI) in routine diagnostic care, assessing its added value against PI-RADS v2.1 and validating results against MRI-guided prostate biopsy.

# 1. Introduction

*1.1. Prostate Cancer Epidemiology*

Prostate cancer (PCa) remains a major global health challenge. The 2022–2027 National Cancer Plan official communication, based on GLOBOCAN data, reported that lung, prostate, and colorectal cancer are the three most common malignancies among men worldwide [1]. PCa is the most frequently diagnosed cancer in men in 118 of 185 countries. Cancer incidence varies by socioeconomic development. Countries undergoing economic transition face rising risks of certain cancers, particularly breast cancer in women and PCa in men. In Latin America, overall cancer mortality is projected to increase by more than 100% in the coming years.

*1.2. Overall Prostate Cancer Diagnostic Workflow*

Risk-adaptive algorithms first combine prostate-specific antigen (PSA), digital rectal exam (DRE), prior biopsy family history and routine pre-biopsy multiparametric MRI (mpMRI). mpMRI classifies lesions with Prostate Imaging–Reporting and Data System (PI-RADS) score form 1-5 to estimate the likelihood of clinically significant prostate cancer (csPCa). If MRI is negative (PI-RADS 1–2) and clinical suspicion is low (e.g., PSA density <0.20 ng/mL/cc; for PI-RADS 3, <0.10), biopsy will be deferred with PSA follow-up; otherwise, targeted and perilesional biopsy should be performed for higher scores (PI-RADS ≥4), in accordance with EAU–EANM–ESTRO–ESUR–ISUP–SIOG guidelines [2]. Prostate cancer aggressiveness is graded by Gleason score (Gs) or International Society of Urological Pathology (ISUP) grade groups. A score ≥ 7 Gs or ISUP ≥ 2 signals highly cellular, clinically significant disease, which is more likely to progress and therefore mandates active treatment [3].

*1.3. Relevance of Non-Invasive Evaluation by mpMRI*

Multiparametric MRI has reshaped prostate cancer care by detecting clinically significant disease while reducing unnecessary biopsies and overtreatment of clinically insignificant lesions [4–6]. Standard mpMRI combines T2-weighted, contrast-enhanced and diffusion-weighted imaging, which are scored on the PI-RADS scale to estimate csPCa likelihood and guide biopsy decisions [7,8]. The updated PI-RADS v2.1 further improves tumor detection within the transition zone by 7–18 percentage points [9].

*1.4. Diagnostic Limitations and Inter-Observer Variability of mpMRI*

Despite these advances, mpMRI remains vulnerable to both false negatives and false positives. Inter-observer agreement for PI-RADS is moderate to substantial, with reported κ values of 0.62–0.70 for v2.0 and 0.70–0.72 for v2.1 [10,11]. Moreover, diagnostic accuracy varies widely across studies applying PI-RADS v2.1, with reported sensitivity ranging from 72–96.3%, specificity 62–93.5%, positive predictive value 53–78%, and negative predictive value 64–94.3% [12–15]. This variability hampers reliable interpretation for clinical decision-making. Further development of MRI techniques is needed to improve diagnostic performance, enabling accurate differentiation between benign disease, clinically insignificant prostate cancer (CIS), and csPCa, thereby reducing unnecessary biopsies and supporting cost-effective, imaging-based risk prediction.

*1.5. Time-Dependent Diffusion MRI*

Time-dependent diffusion (TDD) MRI is a relatively new sequence that probes water molecule diffusivity over short diffusion times, potentially revealing tissue microstructure beyond single-cell resolution. Early animal studies demonstrated sensitivity to microscopic tumor features [16–19], but clinical adoption was limited by the need for very high gradient strengths [20,21].

Recent pilot studies in head and neck, breast, and prostate cancers demonstrated technical feasibility even with limited diffusion time ranges [22–25]. Using high-performance gradients ($\leq 80$ mT/m), Xu *et al.* advanced biophysical modeling to extract quantitative microstructural maps from TDD data. However, the approach was not yet feasible for routine diagnostics [23,26]. A breakthrough came when Wu *et al.* showed that standard clinical MRI systems could acquire TDD images that distinguished csPCa from CIS, exploiting the stronger diffusion-time dependence observed in highly cellular, high-grade tumors [27]. This line of research advances the concept of virtual prostate pathology through non-invasive imaging methods.

*1.6. Further Possibilities with the TDD sequence*

Further validation of time-dependent diffusion (TDD) microstructural parameters for PCa imaging must follow a staged evidence framework that confirms: (1) technical adequacy for better lesion detection, (2) diagnostic accuracy, (3) influence on physicians' clinical reasoning, and (4) impact on patient management—culminating in outcome assessment within Fryback & Thornbury's hierarchical model [28,29]. Independent, real-world studies should avoid biases such as the prostatectomy-enriched sample used by Wu *et al.* [27], which may overestimate test performance. Finally, TDD invites development of quantitative MRI metrics, moving beyond simple parameter averaging, to provide objective, low-variability tissue characterization that can inform clinical guidelines.

*1.7. Deep Learning-Based Interpretation*

Accurate lesion localization is essential, as the peripheral and transition zones of the prostate differ markedly in architecture and cancer prevalence [30]. Advances in deep learning (DL) have enabled automated zone segmentation on both mpMRI and transrectal ultrasound, making zone-specific assessment feasible in routine practice [31,32].

Future work should therefore evaluate zone-specific diagnostic performance of TDD-derived microstructural parameters and test machine-learning algorithms for imaging-based prostate cancer risk estimation, particularly given that reproducibility remains hampered by high inter-observer variability [33]. Within the standard diagnostic workflow, novel imaging techniques aim to reduce unnecessary biopsies by offering accurate, non-invasive detection of csPCa [34].

Accordingly, this study will prospectively assess the diagnostic contribution of TDD-derived imaging metrics integrated into a deep learning platform, comparing their performance in predicting csPCa against PI-RADS v2.1–driven mpMRI in real-world clinical practice.

## 2. Materials and Methods

*2.1 Study Design*

This is a prospective, observational, and analytical study, incorporating an exploratory component aimed at developing and validating a novel MRI protocol and AI-based interpretation framework. The study will be conducted at the Clinical Hospital of the University of Chile (HCUCH) and the Faculty of Medicine of the University of Chile. The protocol will be submitted for approval to the ethics committees of both institutions, and research activities will commence only after obtaining formal authorization. No commercial entity has participated in the design of this study, nor will any be involved in its execution or analysis.

*2.2. Study Population*

Consecutive patients with clinical suspicion of PCa—defined as an elevated prostate-specific antigen (PSA) level >4 ng/mL, an abnormal digital rectal examination, or both—without prior prostate biopsy will be recruited from the outpatient clinics. Full eligibility criteria are summarized in **Figure 1**. All participants will provide written informed consent before enrollment.

According to the local standard-of-care diagnostic pathway, men with clinical suspicion of PCa undergo mpMRI (including T2-weighted, diffusion-weighted, and dynamic contrast-enhanced sequences). The dominant lesion is scored by a senior radiologist with at least 10 years of experience, using the PI-RADS v2.1 system:

- PI-RADS 1: very low (clinically significant cancer highly unlikely)
- PI-RADS 2: low (clinically significant cancer unlikely)
- PI-RADS 3: intermediate (equivocal)
- PI-RADS 4: high (clinically significant cancer likely)
- PI-RADS 5: very high (clinically significant cancer highly likely)

Patients with PI-RADS 1–2 findings are not offered biopsy, while those with PI-RADS 3–5 lesions undergo prostate biopsy as per standard-of-care, following the Ginsburg protocol [35]. Biopsy results are interpreted according to the EAU–EANM–ESTRO–ESUR–ISUP–SIOG guidelines [2] and serve as the diagnostic reference standard for this study.

For the purpose of this study, all enrolled patients will undergo an additional 4.5-minute TDD sequence during the same visit as the standard mpMRI. All image data will be exported to a dedicated study database. Only the standard-of-care mpMRI sequences will be stored in the institutional Picture Archiving and Communication Systems (PACS) for clinical use.

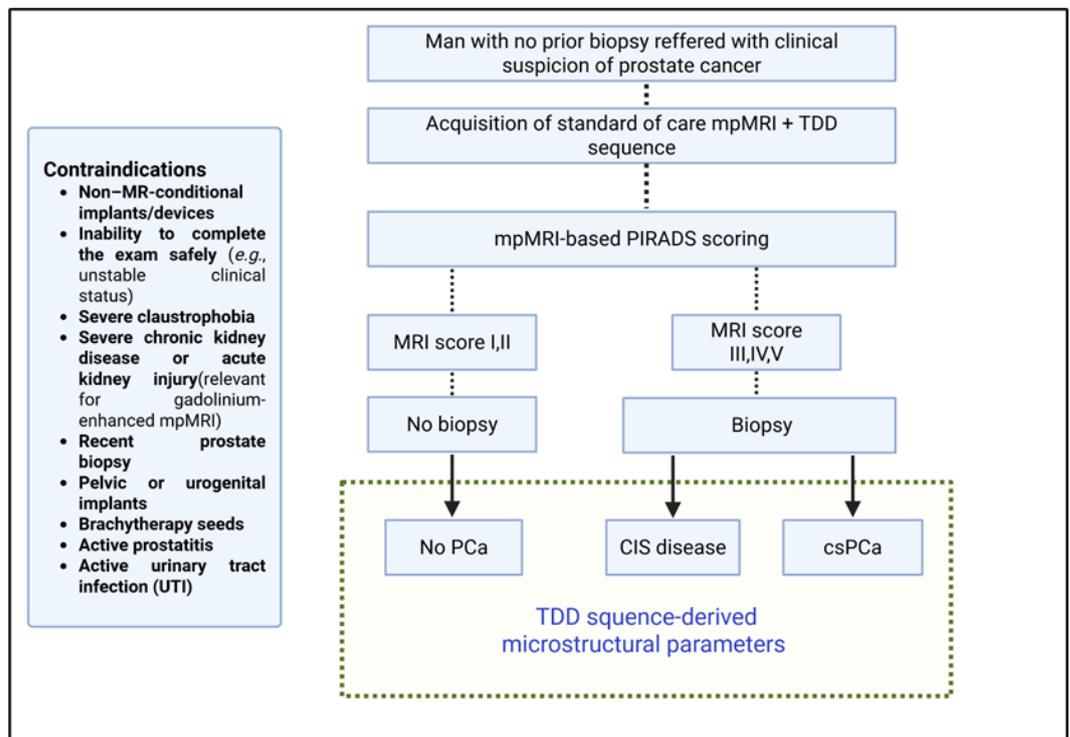

Figure 1. Study flow and imaging protocol. Common MRI/contrast contraindications are summarized at left. TDD-derived microstructural parameters were computed for each outcome group. mpMRI, multiparametric MRI; TDD, temporal diffusion dispersion; PI-RADS, Prostate Imaging–

Reporting and Data System; csPCa, clinically significant prostate cancer; CIS, clinically insignificant prostate cancer.

*2.3. Prostate magnetic resonance imaging data acquisition and fitting for determination of microstructural parameters*

Multiparametric MRI will be performed on a 3.0-T scanner with the participant in the supine position, using a pelvic phased-array coil, with or without an endorectal coil. Standard mpMRI sequences (T2W, DWI, and DCE) will be acquired following consensus guidelines [36].

An additional TDD sequence will be acquired following the protocol proposed by Wu *et al*. [27]. This protocol uses oscillating gradient spin-echo frequencies of 33 Hz (effective diffusion time = 7.5 msec, two cycles, b = 300 and 600 sec/mm2) and 17 Hz (effective diffusion time = 15 msec, one cycle, b = 400, 800, and 1200 sec/mm2) [27], and pulsed gradient spin-echo sequences at diffusion duration and separation of 10 and 30 msec, respectively (effective diffusion time = 26.7 msec, b = 400, 800, and 1200 sec/mm2), with a maximum gradient of 45 mT/m and maximum slew rate of 200 mT/m/msec [27].

Diffusion data will be analyzed with a two-compartment model with impermeable spheres [26], fitted by nonlinear least squares optimization in Matlab (MathWorks). To avoid local minima, the fitting will be repeated 20 times with randomized initialization. The signal model is given by:

$$S = v_{in} \cdot S_{in} \cdot (1 - v_{in}) \cdot e^{-b \cdot D_{ex}}$$

where $v_{in}$ is the intracellular volume fraction, $S_{in}$ the intracellular signal (as defined by Jiang et al. [26] for trapezoidal OGSE encoding [23]), $b$ is the diffusion-weighting factor and $D_{ex}$ the extracellular diffusivity [19,26]. The Matlab routine provided by Jiang et al. [37] will be used.

*2.4. Artificial intelligence-based automatic delineation of the prostate gland zones*

This study aims to develop and validate an AI-based framework for the automatic delineation of prostate zones and suspicious lesions, enabling accurate three-dimensional quantification of microstructural parameters for the prediction of csPCa.

2.4.1. Training Dataset and Human-in-the-Loop Strategy

DL-based segmentation requires annotated data. We will begin with the publicly available PROSTATEx dataset [38] and its accompanying annotations [39]. The relevance of this dataset is that it comprises mpMRI examinations collected for a challenge on prostate lesion detection and classification. It provides T2-weighted (T2W), diffusion-weighted (DWI), and apparent diffusion coefficient (ADC) images acquired on Siemens scanners, along with lesion annotations and clinical significance labels (biopsy-confirmed Gleason Grade Group) for a subset of patients. For local patient data, a Human-in-the-Loop strategy will be implemented: batches of automatic segmentations from a 3D U-Net trained on the public dataset will be iteratively corrected by expert radiologists [40]. Corrected masks are added to the training set and the model is retrained until convergence, defined by radiologist-established criteria. Model performance will be evaluated on an unseen test set using the Dice coefficient or more sophisticated Dice variants measurements.

2.4.2. Segmentation Models

Automatic segmentation will primarily rely on nnU-Net [41], a self-configuring deep learning framework that automatically adapts preprocessing, network architecture, training schedules, and post-processing to the characteristics of each dataset. This approach builds on the U-Net paradigm, employing a contracting path to capture contextual information and an expanding path to delineate regions of interest with high precision. For additional flexibility and integration of custom architectures, we will also leverage the Medical Open Network for Artificial Intelligence (MONAI) framework [42]. MONAI is a PyTorch-based, open-source framework that provides medical imaging–specific transforms, model architectures, and utilities to streamline the development and deployment of clinical AI models. Together, nnU-Net and MONAI establish a robust, state-of-the-art segmentation pipeline that is reproducible and adaptable to diverse imaging datasets.

In addition to these baselines, we will consider foundation models such as MedSAM, which leverage large-scale pretraining for cross-domain adaptability and can be adapted for automatic use via lightweight fine-tuning or prompt generation [43]. MedSAM is trained on over 1.5 million annotated medical images and significantly improves segmentation performance across a wide variety of anatomical structures and imaging modalities. We will also evaluate ProGNet [44], a U-Net derivative optimized for automatic prostate segmentation. ProGNet has been retrospectively validated on 905 patients and prospectively tested, achieving mean Dice coefficients of 0.92 and 0.93, respectively, while reducing segmentation time from >10 minutes manually to ~35 seconds per case. Its code is publicly available for reproducibility and external validation [44,45].

2.4.3. Integration with Microstructural Analysis

Automatic segmentations will define ROIs for TDD-derived microstructural parameter extraction. Combined with biopsy results, these metrics will train the PROSTDAI DL model for lesion-level risk prediction, aiming to improve reproducibility, reduce inter-observer variability, and support clinical decision-making.

*2.5. Radiological imaging analysis*

All standard mpMRI examinations (excluding the TDD sequence) will be interpreted according to PI-RADS v2.1 by radiologists through PACS. After signing the report, the radiologist will segment the most relevant lesion within one month. These segmented regions will define ROIs for extracting microstructural parameters from co-registered TDD images.

Microstructural parameters: intracellular volume fraction ($v_{in}$), cell diameter ($d$), extracellular diffusivity ($D_{ex}$) and cellularity index ($v_{in}/d$), will be estimated using METSC [46], a Transformer-based architecture designed for multi-shell diffusion MRI. METSC addresses limitations of conventional non-linear optimization in biophysical multi-compartment models by learning latent representations directly from q-space samples. This approach reduces the number of diffusion encodings required, demonstrating reduced scan times while preserving or improving fitting accuracy.

TDD-derived microstructural metrics obtained with METSC will be evaluated for their ability to classify lesion categories and will contribute to the training and validation of the PROSTDAI AI framework.

*2.6. Histopathologic analysis*

The interval between MRI examination and prostate biopsy will be limited to 30 days. All biopsy specimens will be interpreted by senior pathologists with over 10 years of experience, following the International Society of Urological Pathology (ISUP) grading standards [47]. The GSs of all tumor foci will be recorded and the findings grouped into five categories, as per the ISUP grade groups (GGs): GS 3+3 (GG 1), GS 3+4 (GG 2), GS 4+3 (GG 3), GS 4+4 (GG 4), and GS over 4+4 (GG 5). For analysis purposes, benign lesions and PCa GG 1 will be classified as clinically insignificant prostate cancer (CIS), while lesions within GGs 2–5 will be classified as csPCa.

*2.7. Statistical analysis*

Microstructural parameters will be measured within the manually delineated lesion ROIs, excluding boundary voxels to minimize partial volume effects. Previous studies have primarily considered the mean values of these parameters [27]; however, it remains unknown whether alternative summary statistics may provide superior discrimination between clinically insignificant (CIS) and clinically significant prostate cancer (csPCa).

Post-hoc comparisons among tissue categories (benign tissue and ISUP grade groups) will be performed using one-way analysis of variance (ANOVA) followed by pairwise t-tests with Tukey correction [48]. All analyses will be conducted using the R Studio and IBM SPSS Statistics software environment.

The diagnostic performance of TDD-derived microstructural parameters in distinguishing csPCa from CIS will be evaluated using linear discriminant analysis (LDA), support vector machine (SVM), Random Forest and extreme gradient boosting (XGBoost), with fivefold cross-validation in Matlab. Classification performance of individual PI-RADS-based scoring and TDD MRI parameters will be assessed using standard metrics: accuracy, sensitivity, specificity, and area under the receiver operating characteristic curve (AUC). A significance level of $P<0.05$ will be considered statistically significant for all tests.

## 3. Expected Results

*3.1. Deep Learning Models*

We anticipate developing two complementary deep learning models for automated prostate analysis:

- Prostate Segmentation Model: Based on U-Net or ProGNet architectures, initially trained on the PROSTATEx dataset [38] and subsequently fine-tuned using multiparametric MRI (mpMRI) data from the Clinical Hospital of the University of Chile (HCUCH). This model is expected to accurately delineate the prostate gland and serve as a pre-processing step for downstream microstructural analysis. For the segmentation task, we anticipate achieving a Dice Similarity Coefficient (DSC) of approximately 0.92, consistent with state-of-the-art literature.
- Tissue Microstructure Estimation Model: Employing a Transformer-based architecture inspired by sparse representation techniques; METSC [46], this model will estimate voxel-wise tissue microstructural parameters, including intracellular and extracellular volume fractions and diffusivities, from multi-shell diffusion MRI (dMRI) data. These microstructural parameters will then be used for tissue classification. For microstructure estimation and lesion classification, we anticipate an accuracy exceeding 80%.

Together, these models will form a fully automated pipeline integrating anatomical and microstructural information to support clinical decision-making in PCa diagnostics.

*3.2. Pipeline Integration*

We propose an end-to-end deep learning pipeline that combines anatomical segmentation with microstructure-based tissue classification. The pipeline will be trained end-to-end, by a deep neural network that combines imaging data with clinical features, thereby simultaneously optimizing segmentation and classification objectives. This joint learning approach enables shared feature representations, enhancing both anatomical delineation and microstructural characterization. By integrating diffusion-informed biomarkers with anatomical information, the framework is expected to improve diagnostic accuracy and provide clinically meaningful predictions.

## 4. Discussion

The transformative potential of this project lies in the development and evaluation of novel imaging techniques and analytic approaches aimed at improving the non-invasive detection of csPCa. By integrating time-dependent diffusion MRI (TDD-MRI) with AI-based quantitative analysis, this study seeks to enhance diagnostic specificity, reduce the reliance on invasive biopsy procedures, and provide accurate, low-risk, and more accessible risk stratification for men at intermediate clinical risk.

Demonstrating higher specificity and reliable performance of these novel imaging methods could have a significant impact on both research and clinical practice. Positive findings would provide the foundation for future randomized controlled trials and support evidence-based updates to current standard-of-care diagnostic workflows, potentially leading to more personalized and efficient patient management.

This project emphasizes the translational value of basic and preclinical research into clinically actionable applications. By focusing on clinically meaningful endpoints, it bridges the gap between advanced imaging research and real-world patient care, supporting evidence-driven decision-making in prostate cancer management.

## 5. Conclusions

This section is not mandatory but can be added to the manuscript if the discussion is unusually long or complex.

## 6. Patents

This section is not mandatory but may be added if there are patents resulting from the work reported in this manuscript.

**Author Contributions:** Conceptualization, C.G.S.; methodology, C.G.S., G.C., C.A., V.C., I.G. and C.G.; software, V.C. and C.G.; validation, G.C., C.A., I.G., R.S. and P.N.; investigation, C.G.S. and B.R.; writing—original draft preparation, C.G.S., S.G.C., V.C. and B.R.; writing—review and editing, C.G., P.N., R.S., I.G., C.A. and G.C.; visualization, C.G.S. and B.R.; supervision, C.G.S., G.C., C.A., V.C., I.G. and C.G.; project administration, C.G.S. All authors have read and agreed to the published version of the manuscript. Please turn to the [CRediT taxonomy](#) for the term explanation.

**Funding:** This research received no external funding.

**Institutional Review Board Statement:** The study was conducted in accordance with the Declaration of Helsinki, and approved by the Institutional Review Board (or Ethics Committee) of the University of Chile Clinical Hospital (protocol code 1339/23, approved on may 9th, 2023).

**Informed Consent Statement:** Written informed consent will be obtained from all patients prior to trial participation.

**Data Availability Statement:** The original contributions presented in this study are included in the article. Further inquiries can be directed to the corresponding author.

**Acknowledgments:** The authors thank José de Grazia, Gonzalo Miranda, Jorge Díaz-Jara and Patricio Palavecino for their collaboration and contribution to performing this study.

**Conflicts of Interest:** The authors declare no conflicts of interest.

## Abbreviations

The following abbreviations are used in this manuscript:

PCa  Prostate cancer
PSA  Prostate-specific antigen

| | |
|---|---|
| DRE | Digital rectal exam |
| mpMRI | multiparametric MRI |
| csPCa | Clinically significant prostate cancer |
| Gs | Gleason score |
| ISUP | International Society of Urological Pathology |
| CIS | Clinical insignificant prostate cancer |
| TDD | Time-dependent diffusion |
| GG | Gleason grade |

## References

1. International Agency for Research on Cancer. *World: Cancer fact sheet (GLOBOCAN 2022, version 1.1)* [Internet]. Lyon (FR): International Agency for Research on Cancer; 2024 [cited 2025 Jul 23]. Available from: https://gco.iarc.who.int/media/globocan/fact-sheets/populations/900-world-fact-sheet.pdf
2. Cornford P, van den Bergh RCN, Briers E, Van den Broeck T, Brunckhorst O, Darraugh J, et al. EAU-EANM-ESTRO-ESUR-ISUP-SIOG Guidelines on Prostate Cancer – 2024 Update. Part I: Screening, Diagnosis, and Local Treatment with Curative Intent. *Eur Urol*. 2024;86(2):148-163. doi:10.1016/j.eururo.2024.03.027
3. Chatterjee A, Watson G, Myint E, Sved P, McEntee M, Bourne R. Changes in epithelium, stroma, and lumen space correlate more strongly with Gleason pattern and are stronger predictors of prostate ADC changes than cellularity metrics. *Radiology*. 2015;277(3):751-62. doi:10.1148/radiol.2015142414
4. Mortezavi A, Märzendorfer O, Donati OF, Rizzi G, Rupp NJ, Wettstein MS, et al. Diagnostic accuracy of multiparametric magnetic resonance imaging and fusion-guided targeted biopsy evaluated by transperineal template saturation prostate biopsy for the detection and characterization of prostate cancer. *J Urol*. 2018;200(2):309-18. doi:10.1016/j.juro.2018.02.067
5. van der Leest MMG, Cornel EB, Israël B, Hendriks R, Padhani AR, Hoogenboom M, et al. Head-to-head comparison of transrectal ultrasound-guided prostate biopsy versus multiparametric prostate resonance imaging with subsequent magnetic resonance-guided biopsy in biopsy-naïve men with elevated prostate-specific antigen: a large prospective multicenter clinical study. *Eur Urol*. 2019;75(4):570-8. doi:10.1016/j.eururo.2018.11.023
6. Ahmed HU, El-Shater Bosaily A, Brown LC, Gabe R, Kaplan R, Parmar MK, et al. Diagnostic accuracy of multi-parametric MRI and TRUS biopsy in prostate cancer (PROMIS): a paired validating confirmatory study. *Lancet*. 2017;389(10071):815-22. doi:10.1016/S0140-6736(16)32401-1
7. Pourvaziri A. Mass enhancement pattern at prostate MRI as a potential PI-RADS criterion. Radiol Imaging Cancer. 2022;4(2):e229006. doi:10.1148/rycan.229006
8. Park SY, Park BK, Kwon GY. Diagnostic performance of mass enhancement on dynamic contrast-enhanced MRI for predicting clinically significant peripheral zone prostate cancer. AJR Am J Roentgenol. 2020;214(4):792-799. doi:10.2214/AJR.19.22072
9. Agrotis G, Pais Pooch E, Marsitopoulos K, Vlychou M, Benndorf M, Beets-Tan RGH, Schoots IG. Detection rates for prostate cancer using PI-RADS 2.1 upgrading rules in transition zone lesions align with risk assessment categories: a systematic review and meta-analysis. Eur Radiol. 2025 Apr 27. doi:10.1007/s00330-025-11618-w
10. Wei CG, Zhang YY, Pan P, Chen T, Yu HC, Dai GC, et al. Diagnostic accuracy and interobserver agreement of PI-RADS Version 2 and Version 2.1 for the detection of transition zone prostate cancers. *AJR Am J Roentgenol*. 2021;216(5):1247-56. doi:10.2214/AJR.20.23883
11. Bhayana R, O'Shea A, Anderson MA, Bradley WR, Gottumukkala RV, Mojtahed A, et al. PI-RADS Versions 2 and 2.1: Interobserver agreement and diagnostic performance in peripheral and transition zone lesions among six radiologists. *AJR Am J Roentgenol*. 2021;217(1):141-51. doi:10.2214/AJR.20.24199
12. Wen J, Liu W, Shen X, Hu W. PI-RADS v2.1 and PSAD for the prediction of clinically significant prostate cancer among patients with PSA levels of 4–10 ng/ml. *Sci Rep*. 2024;14:6570. doi:10.1038/s41598-024-57337-y
13. Lee CH, Vellayappan B, Tan CH. Comparison of diagnostic performance and inter-reader agreement between PI-RADS v2.1 and PI-RADS v2: systematic review and meta-analysis. *Br J Radiol*. 2022;95(1131):20210509. doi:10.1259/bjr.20210509
14. Wang W, Zhu M, Luo Z, Li F, Wan C, Zhu L. Diagnostic value analysis of PI-RADS v2.1 combined with ADC values in the risk stratification of prostate cancer Gleason scores: a retrospective study. *Arch Esp Urol*. 2024;77(8):889-896. doi:10.56434/j.arch.esp.urol.20247708.125.
15. Wei X, Xu J, Zhong S, Zou J, Cheng Z, Ding Z, *et al*. Diagnostic value of combining PI-RADS v2.1 with PSAD in clinically significant prostate cancer. *Abdom Radiol (NY)*. 2022;47(10):3574-3582. doi:10.1007/s00261-022-03592-4

16. Colvin DC, Yankeelov TE, Does MD, Yue Z, Quarles C, Gore JC. New insights into tumor microstructure using temporal diffusion spectroscopy. *Cancer Res*. 2008;68(14):5941-7. doi:10.1158/0008-5472.CAN-08-0832
17. Colvin DC, Loveless ME, Does MD, Yue Z, Yankeelov TE, Gore JC. Earlier detection of tumor treatment response using magnetic resonance diffusion imaging with oscillating gradients. *Magn Reson Imaging*. 2011;29(3):315-23. doi:10.1016/j.mri.2010.10.003
18. Xu J, Li K, Smith RA, Waterton JC, Zhao P, Chen H, et al. Characterizing tumor response to chemotherapy at various length scales using temporal diffusion spectroscopy. *PLOS One*. 2012;7(7):e41714. doi:10.1371/journal.pone.0041714
19. Reynaud O, Winters KV, Hoang DM, Wadghiri YZ, Novikov DS, Kim SG. Surface-to-volume ratio mapping of tumor microstructure using oscillating gradient diffusion-weighted imaging. *Magn Reson Med*. 2016;76(1):237-47. doi:10.1002/mrm.25865
20. Baron CA, Beaulieu C. Oscillating gradient spin-echo (OGSE) diffusion tensor imaging of the human brain. *Magn Reson Med*. 2014;72(3):726-736. doi:10.1002/mrm.24987
21. Van AT, Holdsworth SJ, Bammer R. In vivo investigation of restricted diffusion in the human brain with optimized oscillating diffusion gradient encoding. *Magn Reson Med*. 2014;71(1):83-94. doi:10.1002/mrm.24632
22. Iima M, Yamamoto A, Kataoka M, Yamada Y, Omori K, Feiweier T, et al. Time-dependent diffusion MRI to distinguish malignant from benign head and neck tumors. *J Magn Reson Imaging*. 2019;50(1):88-95. doi:10.1002/jmri.26578
23. Xu J, Jiang X, Li H, Arlinghaus LR, McKinley ET, Devan SP, et al. Magnetic resonance imaging of mean cell size in human breast tumors. *Magn Reson Med*. 2020;83(6):2002-14. doi:10.1002/mrm.28056
24. Lemberskiy G, Rosenkrantz AB, Veraart J, Taneja SS, Novikov DS, Fieremans E. Time-dependent diffusion in prostate cancer. *Invest Radiol*. 2017;52(7):405-11. doi:10.1097/RLI.0000000000000356
25. Xu J, Jiang X, Devan SP, Arlinghaus LR, McKinley ET, Xie J, et al. MRI-cytometry: Mapping non-parametric cell size distributions using diffusion MRI. *Magn Reson Med*. 2021;85(2):748-61. doi:10.1002/mrm.28454
26. Jiang X, Li H, Xie J, McKinley ET, Zhao P, Gore JC, et al. In vivo imaging of cancer cell size and cellularity using temporal diffusion spectroscopy. *Magn Reson Med*. 2017;78(1):156-164. doi:10.1002/mrm.26356
27. Wu D, Jiang K, Li H, Zhang Z, Ba R, Zhang Y, et al. Time-dependent diffusion MRI for quantitative microstructural mapping of prostate cancer. *Radiology*. 2022;303(3):578-87. doi:10.1148/radiol.211180
28. Chatterjee A, Oto A. Prostate tissue microstructural estimates using time-dependent diffusion MRI. Radiology. 2022;303(3):588-589. doi:10.1148/radiol.220056
29. Fryback DG, Thornbury JR. The efficacy of diagnostic imaging. Med Decis Making. 1991;11(2):88-94. doi:10.1177/0272989X9101100203
30. Purysko AS, Rosenkrantz AB, Barentsz JO, Weinreb JC, Macura KJ. PI-RADS Version 2: A pictorial update. *Radiographics*. 2016;36(5):1354-1372. doi:10.1148/rg.2016150234
31. van Sloun RJG, Wildeboer RR, Mannaerts CK, Postema AW, Gayet M, Beerlage HP, et al. Deep learning for real-time, automatic, and scanner-adapted prostate (zone) segmentation of transrectal ultrasound, for example, magnetic resonance imaging–transrectal ultrasound fusion prostate biopsy. *Eur Urol Focus*. 2021;7(1):78-85. doi:10.1016/j.euf.2019.04.009
32. Bardis M, Houshyar R, Chantaduly C, Tran-Harding K, Ushinsky A, Chahine C, et al. Segmentation of the prostate transition zone and peripheral zone on MR images with deep learning. *Radiol Imaging Cancer*. 2021;3(3):e200024. doi:10.1148/rycan.2021200024
33. Glazer DI, Mayo-Smith WW, Sainani NI, Sadow CA, Vangel MG, Tempany CM, et al. Interreader agreement of Prostate Imaging Reporting and Data System version 2 using an in-bore MRI-guided prostate biopsy cohort: a single institution's initial experience. *AJR Am J Roentgenol*. 2017;209(3):W145-W151. doi:10.2214/AJR.16.17551
34. Hectors SJ, Chen C, Chen J, Wang J, Gordon S, Yu M, et al. Magnetic resonance imaging radiomics-based machine learning prediction of clinically significant prostate cancer in equivocal PI-RADS 3 lesions. *J Magn Reson Imaging*. 2021;54(5):1466-1473. doi:10.1002/jmri.27692
35. Hansen NL, Patruno G, Wadhwa K, Gaziev G, Miano R, Barrett T, et al. Magnetic resonance and ultrasound image fusion supported transperineal prostate biopsy using the Ginsburg protocol: technique, learning points, and biopsy results. *Eur Urol*. 2016;70(2):332-340. doi:10.1016/j.eururo.2016.02.064
36. de Rooij M, Israël B, Tummers M, *et al*. ESUR/ESUI consensus statements on mpMRI quality requirements. *Eur Radiol*. 2020;30(10):5404-5416. doi:10.1007/s00330-020-06929-z
37. Emoryzzl. Prostate_impulse_MRI [Internet]. GitHub; [cited 2025 Aug 26]. Available from: https://github.com/Emoryzzl/Prostate_impulse_MRI

38. Litjens G, Debats O, Barentsz J, Karssemeijer N, Huisman H. SPIE-AAPM PROSTATEx Challenge Data (Version 2) [dataset]. The Cancer Imaging Archive; 2017. doi:10.7937/K9TCIA.2017.MURS5CL
39. Cuocolo R, Stanzione A, Castaldo A, De Lucia DR, Imbriaco M. Quality control and whole-gland, zonal and lesion annotations for the PROSTATEx challenge public dataset. **Eur J Radiol.** 2021;138:109647. doi:10.1016/j.ejrad.2021.109647
40. Vásquez-Venegas C, Sotomayor CG, Ramos B, Castañeda V, Pereira G, Cabrera-Vives G, Härtel S. Human-in-the-Loop — A Deep Learning Strategy in Combination with a Patient-Specific Gaussian Mixture Model Leads to the Fast Characterization of Volumetric Ground-Glass Opacity and Consolidation in the Computed Tomography Scans of COVID-19 Patients. J Clin Med. 2024;13(17):5231. doi:10.3390/jcm13175231
41. Isensee F, Jaeger PF, Kohl SAA, Petersen J, Maier-Hein KH. nnU-Net: Self-adapting framework for U-Net-based medical image segmentation [Internet]. GitHub; 2025 [cited 2025 Sep 28]. Available from: https://github.com/MIC-DKFZ/nnUNet
42. Project MONAI. MONAI—Medical Open Network for AI [Internet]. [cited 2025 Aug 26]. Available from: https://monai.io
43. Ma J, Wang B, An L, Zhang Y, Wang H, et al. MedSAM: Segment Anything in Medical Images [Internet]. GitHub; 2025 [cited 2025 Sep 28]. Available from: https://github.com/bowang-lab/MedSAM
44. Soerensen SJC. ProGNet: Deep learning for automatic prostate segmentation [Internet]. GitHub; 2021 [cited 2025 Sep 28]. Available from: https://github.com/simonjcs/ProGNet
45. Soerensen SJC, Fan RE, Seetharaman A, Chen L, Shao W, Bhattacharya I, Kim YH, Sood R, Borre M, Chung BI, To'o KJ, Rusu M, Sonn GA. Deep learning improves speed and accuracy of prostate gland segmentations on magnetic resonance imaging for targeted biopsy. **J Urol.** 2021;206(3):604-612. doi:10.1097/JU.0000000000001783
46. Zheng T, Yan G, Li H, Zheng W, Shi W, Zhang Y, et al. A microstructure estimation Transformer inspired by sparse representation for diffusion MRI. **Med Image Anal.** 2023;86:102788. doi:10.1016/j.media.2023.102788
47. Epstein JI, Egevad L, Amin MB, Delahunt B, Srigley JR, Humphrey PA; Grading Committee. The 2014 International Society of Urological Pathology (ISUP) Consensus Conference on Gleason Grading of Prostatic Carcinoma: definition of grading patterns and proposal for a new grading system. **Am J Surg Pathol.** 2016;40(2):244–252. doi:10.1097/PAS.0000000000000530
48. Tukey JW. Comparing individual means in the analysis of variance. **Biometrics**. 1949;5(2):99–114. doi:10.2307/3001913